\documentclass[twocolumn,secnumarabic,amssymb, nobibnotes, aps, prd]{revtex4-2}

\usepackage{graphicx}
\graphicspath{{./picture/}}
\usepackage{amsmath,amssymb}
\usepackage{mathtools}
\usepackage{physics}
\usepackage{color}
\usepackage{upgreek}
\usepackage{hyperref}  
\usepackage{bm}
\newcommand{\micron}{{\upmu\mathrm{m}}}

\newcommand{\rc}[1]{\textcolor{black}{#1}}
\newcommand{\RC}[1]{\textcolor{black}{#1}}

\begin{document}

\title{Positron generation and acceleration in a self-organized photon collider\\ enabled by an ultra-intense laser pulse}%

\author{K. Sugimoto$^{1,2}$, Y. He$^{3}$, N. Iwata$^{2,4}$, I-L. Yeh$^{5}$, K. Tangtartharakul$^{3}$, A. Arefiev$^{3}$, and Y. Sentoku$^{2}$}


\affiliation{$^{1}$Department of Physics, Graduate School of Science, Osaka University, 1-1 Machikanecho, Toyonaka, Osaka 560-0043, Japan}
\affiliation{$^{2}$Institute of Laser Engineering, Osaka University, 2-6 Yamadaoka, Suita, Osaka 565-0871, Japan}
\affiliation{$^{3}$Department of Mechanical and Aerospace Engineering, University of California at San Diego, La Jolla, CA 92093, United States of America}
\affiliation{$^{4}$Institute for Advanced Co-Creation Studies, Osaka University, 1-1 Yamadaoka, Suita, Osaka 565-0871, Japan}
\affiliation{$^{5}$Department of Physics, University of California at San Diego, La Jolla, CA 92093, United States of America}

\date{December 2022}%

\begin{abstract}
\RC{We discovered a simple regime where a near-critical plasma irradiated by a laser of experimentally available intensity can self-organize to produce positrons and accelerate them to ultra-relativistic energies. The laser pulse piles up electrons at its leading edge, producing a strong longitudinal plasma electric field. The field creates a moving gamma-ray collider that generates positrons via the linear Breit-Wheeler process -- annihilation of two gamma-rays into an electron-positron pair.  At the same time, the plasma field, rather than the laser, serves as an accelerator for the positrons. The discovery of positron acceleration was enabled by a first-of-its-kind kinetic simulation that generates pairs via photon-photon collisions. Using available laser intensities of $10^{22}$\,$\rm W/cm^2$, the discovered regime can generate a GeV positron beam with divergence angle of $\sim10^{\circ}$ and total charge of 0.1\,pC. The result paves the way to experimental observation of the linear Breit-Wheeler process and to applications requiring positron beams.}


\end{abstract}

\maketitle

In astrophysics, creation of matter from light is ubiquitous, playing an important role for various astrophysical objects~(e.g. see \cite{medin.2010, Beloborodov_2008, Philippov_2018, uzdensky.2022, Hakobyan_2023}). The advent of ultra-high-intensity laser facilities~\cite{danson2019petawatt_review, ELI_NP_Tanaka, Gist_multi-PW_10_22} promises to enable, for the first time, creation of electron-positrons pairs from light alone on a macroscopic scale in laboratory. If successfully implemented, this capability will open a new area of QED research~\cite{RevModPhys.84.1177,RevModPhys.94.045001,zhang.pop.2020} and it will enable laboratory studies of astrophysically relevant electron-positron plasmas~\cite{MP3}. The ability to generate positrons by a laser is also likely to impact the research on laser-driven positron acceleration. Currently, positrons are produced by an external source and the focus is on finding augmented configurations that facilitate positron acceleration \cite{Gessner_ncomm_2016,Zhou_prl_2021,Silva_prl_2021,Vieira_prl_2014}.

In the context of pair production from light alone, it is important to distinguish between the non-linear~\cite{Burke_prl_1997} and linear~\cite{Breit_pr_1934} Breit-Wheeler (BW) processes. The nonlinear BW or the multiphoton process is the decay of a $\gamma$-ray propagating through a laser pulse into a pair. The decay involves multiple coherent optical photons. The linear BW or the two-photon process is the annihilation of two energetic $\gamma$-rays that leads to pair production. The setups that many pairs via the nonlinear BW~\cite{Ridgers_2012, Vranic_sci_2018, Xing-Long_mre_2019, Jian-Xun_ppcf_2019, Zhao_nat_2022,Mercuri-Baron_2021,martinez.prab.2023} require a laser intensity in excess of $10^{23}$\,$\rm W/cm^2$. The two-photon process has no laser intensity requirement, but it does require a dense population of energetic $\gamma$-rays to overcome the smallness of the cross-section, \rc{$\sigma_{\gamma \gamma} \sim 10^{-25}$~cm$^2$}, and the energy threshold. \RC{A laser-irradiated plasma can  efficiently generate a $\gamma$-ray beam~\cite{nakamura.prl.2012, ji.prl.2014, Stark_prl_2016}, so colliding in vacuum two such beams (produced by two different laser) is a possible approach to produce pairs~\cite{Ribeyre_pre_2016,wang.PhysRevApp.2020}. The inherent $\gamma$-ray beam divergence requires the targets generating $\gamma$-rays to be close to each other and makes experimental implementation challenging. A conceptually different approach is to generate and collide $\gamma$-ray beams inside one target~\cite{Yutong_nat_2021}. It not only allows to overcome the divergence and thus boost the pair yield~\cite{Yutong_nat_2021}, but, more importantly, it offers an unexplored opportunity to accelerate the linear BW positrons. If the positrons can be accelerated and collimated, then this would facilitate their detection, making a first laboratory observation of the linear BW process possible, and enable their use for applications like positron annihilation lifetime spectroscopy \cite{audet.prab.2021,krause.1999}.}

\RC{In this Letter, we present a simple but previously unknown regime where a dense plasma irradiated by a laser of experimentally achievable intensity self-organizes to produce positrons from light alone and accelerate them to ultra-relativistic energies. The laser pulse piles up electrons at its leading edge, producing a strong longitudinal plasma electric field that moves with the pulse. The field creates a moving $\gamma$-ray collider that generates positrons via the linear BW process and, at the same time, serves as an accelerator for the produced positrons. The discovery of the new positron acceleration mechanism and the synergistic interplay between the photon collider and the plasma accelerator was enabled by a first-of-its-kind kinetic simulation that generates pairs via photon-photon collisions. This work builds on an important observation based on post-processed photon data that a single laser-pulse can generate a colliding population of $\gamma$-rays in a dense structured plasma~\cite{Yutong_njp_2021}. We find that the linear BW process produces about $10^7$ pairs at $3 \times 10^{22}$\,$\rm W/cm^2$, whereas the nonlinear BW process produces no pairs at all. About 10\% of the positrons experience the forward acceleration and form a GeV beam with a divergence angle of $10^{\circ}$. The advantage of our regime is that it uses a simple setup and requires only a single laser with intensity already accessible at ELI~\cite{ELI} and CoReLS~\cite{CoReLS}. }

The laser-plasma interaction is \RC{self-consistently} simulated in 2D-3V with the PIC code PICLS~\textcolor{black}{\cite{Sentoku_jcp_2008}} that includes a radiation transport module~\cite{Sentoku_pre_2014,Royle_pre_2017} for energetic photons emitted via synchrotron radiation~\cite{Pandit_pop_2012} and Bremsstrahlung~\cite{Sentoku_pop_1998}. \RC{We have developed a module for simulating the linear BW process [see Supplemental Materials], making PICLS the first PIC code capable of generating linear BW pairs during the laser-plasma interaction and thus suitable for studies of positron dynamics.} In our setup, a 25~fs, $3 \times 10^{22}$~W/cm$^{2}$ laser pulse irradiates a dense uniform carbon plasma (see Supplemental Materials for simulation parameters). {We normalize all electric fields, $\bm{E}$, and use a dimensionless quantity $\bm{a} = |e| \bm{E}/m_{e} c \omega_0$ instead}, where $e$ and $m_e$ are the electron charge and mass, $c$ is the speed of light, and $\omega_{0}$ is the laser frequency corresponding to vacuum wavelength $\lambda = 0.8~\micron$. The laser amplitude is $a_L = 120$. This laser makes electrons ultra-relativistic and renders a plasma with electron density $n_e$ less than $\gamma_L n_c \sim a_L n_c$ transparent, where $\gamma_L\equiv \sqrt{1+a_L^2/2}$ is the electron Lorentz factor for ponderomotive energy~\cite{Wilks_prl_1992} and $n_c = m_e \omega_0^2 / 4\pi e^2$ is the classical critical density. In our main simulation, the initial electron density is $n_{e0} = 2.8 n_c \ll a_L n_c$, so the laser easily propagates into the plasma.


\begin{figure}[hbb]
 \centering
 \includegraphics[width=\linewidth]{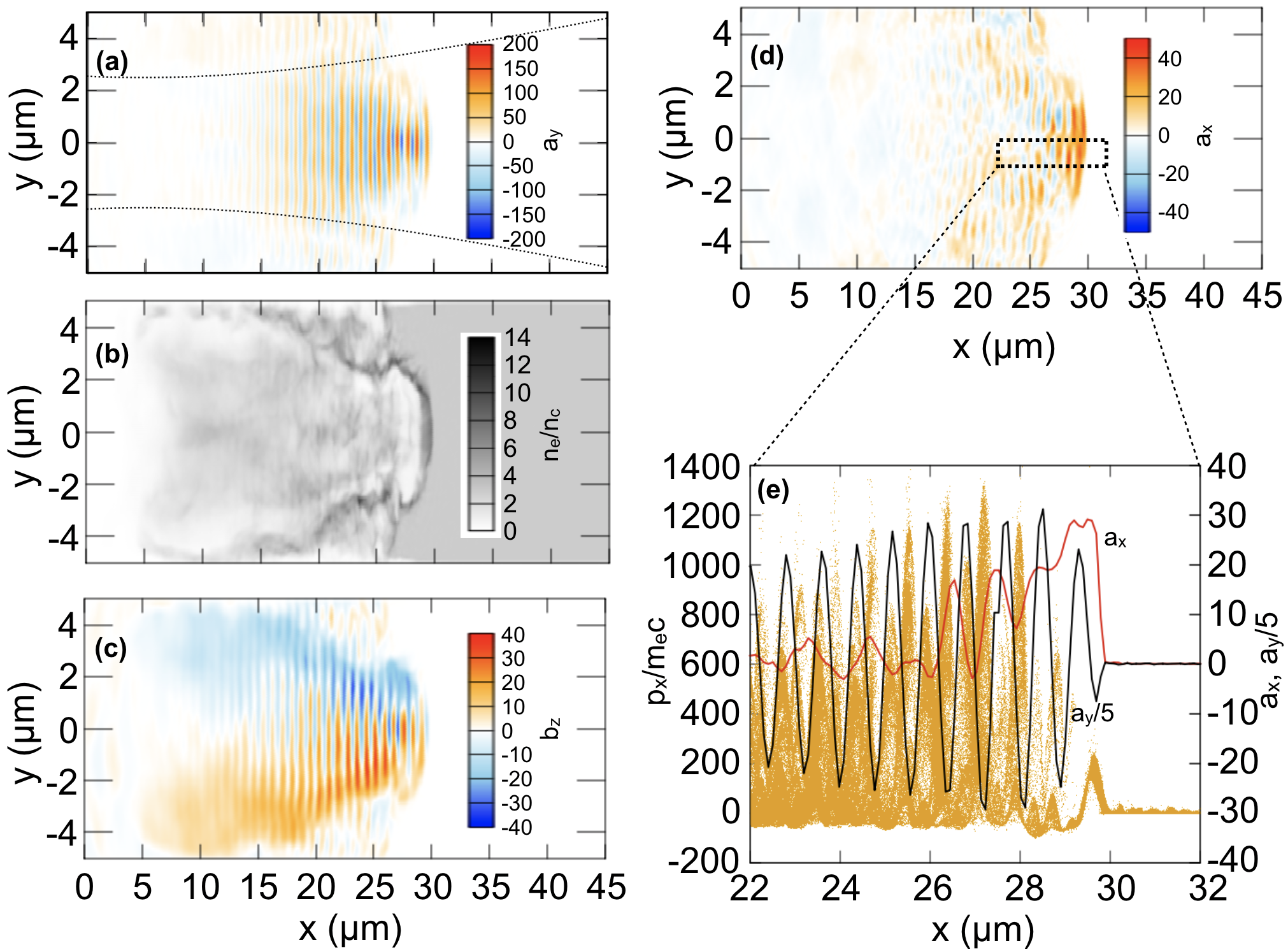}
 \caption{Laser interaction with a dense plasma. (a) Normalized transverse electric field $a_{y}$. Dashed lines indicate the beam waist in the absence of the plasma. (b) Electron density. (c) Normalized magnetic field $b_z$ averaged over one laser period. (d) Normalized longitudinal electric field $a_x$. (e) Electron distribution in the $x$-$p_x$ plane, and $a_x$ and $a_y$ in the vicinity of the pulse front [dashed rectangle in (d)]. The electric fields $a_x$ and $a_y$ in (e) are averaged over $|y| \le 0.5~\micron$. The snapshots in (a)~-~(e) are taken at $t=117$\,fs.}
 \label{fig1}
\end{figure}  

Figure~\ref{fig1} illustrates key aspects of the laser-plasma interaction. All snapshots are taken when the laser pulse reaches $x = 30~\micron$. The corresponding time is $t=117$\,fs, with $t = 0$\,fs being the time when the pulse reaches the target. Figure~\ref{fig1}(a) shows the normalized transverse electric field $a_y$ that is dominated by the field of the laser. 
Due to the relativistic self-focusing, the beam remains tightly focused after having traveled a distance greater than the Rayleigh length ($l_{R}=\pi w_{0}^{2}/\lambda \simeq 25~\micron$ for a focal spot with radius $w_0=2.5~\micron$). The dashed curves mark the expected beam waist in the absence of the target. The self-focusing also increases the laser amplitude to $a_y=150$. The beam becomes fully depleted after propagates $70~\micron$ into the plasma. The profiles of electron density and generated azimuthal magnetic field are shown in Figs.~\ref{fig1}(b)\&(c). Transverse electron expulsion by the ponderomotive force produces a density pileup ($n_e \sim 10 n_c$) at the periphery of the beam that helps guide the laser. The electrons remaining in the beam accelerate forward in the laser field and form longitudinal current. The current generates a strong quasi-static magnetic field $B_z$~\cite{Stark_prl_2016} whose peak strength is 30\% of that for the laser magnetic field. Figure~\ref{fig1}(c) shows the field profile while providing an additional figure of merit $b_{z} = \omega_c / \omega_0$, where $\omega_c = |e| B_{z} / m_e c$ is the cyclotron frequency. 

\begin{figure}[htb]
 \centering
 \includegraphics[width=\linewidth]{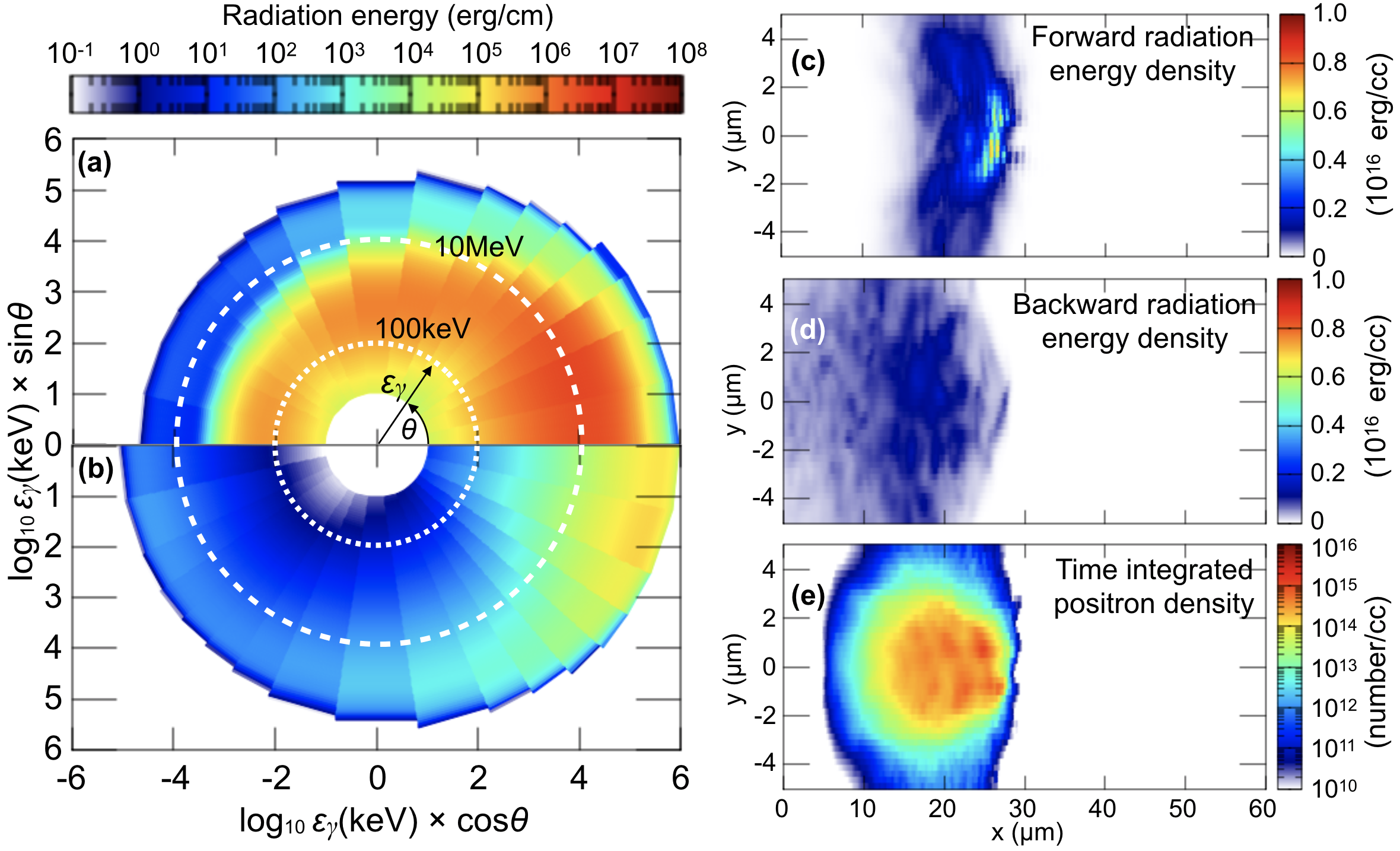}
 \caption{Self-organized photon collider. (a)\&(b) Angular distribution of synchrotron and Bremsstrahlung photons in the region with $22~\micron \le x \le 27.5 \micron$ and $|y| \le 1~\micron$. The radius is $\log_{10} \epsilon_{\gamma} {\rm [keV]}$. The dashed circles are $\epsilon_{\gamma} = 100$\,keV and $10$\,MeV. (c)\&(d) Energy density of forward and backward emitted photons via synchrotron emission. (e) Time integrated number density of the linear BW pair production events. The snapshots in (a)~-~(e) are taken at $t=117$\,fs.}
 \label{fig2}
\end{figure} 

The B-field plays a key role in generating forward-directed $\gamma$-rays. It transversely confines the electrons that are accelerated and pushed forwarded by the laser. \RC{The B-field defects electrons forward instead of causing the conventional rotation} and the deflections change the orientation of the transverse velocity $v_{\perp}$ with respect to $E_{\perp}$ of the laser. If their frequency is comparable to the Doppler-shifted frequency of the laser, then $v_{\perp}$ remains antiparallel to $E_{\perp}$ as the laser field and the electron oscillate. This mechanism of direct laser acceleration assisted by the plasma B-field~\cite{gong.PRE.2020} produces $\sim 500$\,MeV electrons with a forward momentum of 1000\,$m_e c$. They are located in Fig.~\ref{fig1}(e) at $22~\micron \leq x \leq 28~\micron$. The deflections of the electrons by the magnetic field has another important effect -- they cause the electrons to emit MeV $\gamma$-rays in the direction of laser propagation~\cite{Stark_prl_2016, Jansen_ppcf_2018, wang.PhysRevApp.2020}. 

Due to the high plasma density, the laser also generates a strong longitudinal \RC{plasma} electric field that is essential for the production of backward-directed $\gamma$-rays. This is a charge-separation field that arises as the leading edge of the laser pulse sweeps up plasma electrons. \RC{Its peak amplitude is 25\% of $a_y$ and it dominates over the oscillating longitudinal field of the laser.} The positive plasma field is clearly visible in Fig.~\ref{fig1}(d) at $x \approx 29.5~\micron$. After initial forward acceleration to $p_x\sim 200\,m_e c$, the electrons swept up by the leading edge of laser pulse slow down under the influence of $a_x$ 
and then re-accelerate in the backward direction to $p_x\sim -100\,m_ec$. \rc{These electrons emit backward-directed photons. In contrast to the forward-moving electrons, the emission is induced by the laser field~\cite{Koga_PRE_2004} that is much stronger than the plasma magnetic field. This makes the emission more efficient, causing the electrons to quickly lose a large portion of their energy, as seen in Fig.\ \ref{fig1}(e) at $x > 28~\micron$. The emission process accompanies laser propagation since the population of backward-moving electrons is constantly replenished by $a_x$ that is moving forward with the laser pulse.}

The two photon populations form a moving $\gamma$-ray collider. 
Figures~\ref{fig2}(a)\&(b) show photon spectra versus the polar angle $\theta$ 
in the region where the energy density of forward-  and backward-moving photons ($|\theta| \leq \pi/2$ and $|\theta| > \pi/2$) overlap ($22~\micron \le x \le 27.5~\micron$; $|y| \le 1~\micron$). The corresponding energy density plots are shown in Figs.~\ref{fig2}(c)\&(d). The Bremsstrahlung that plays a secondary role is included for completeness. The synchrotron emission converts 40\% of the laser energy into photons over the entire simulation (vs. 2\% for Bremsstrahlung). The linear BW process has a threshold of $\epsilon_{\gamma 1} \epsilon_{\gamma 2} > m_e^2 c^4 \approx 0.26~\mbox{MeV}^2$, where $\epsilon_{\gamma 1,2}$ are the energies of colliding photons. Therefore, linear BW pairs are mainly produced by forward-moving photons with $0.5~\mbox{MeV} \lesssim \epsilon_{\gamma} \lesssim 100~\mbox{MeV}$ colliding with backward-moving photons with $10~\mbox{keV} \lesssim \epsilon_{\gamma} \lesssim 1~\mbox{MeV}$. The photon densities in these two groups are comparable, with $n_{\gamma} \sim 10^{22}$\,cm$^{-3}$.
The probability for a backward-moving photon to produce a pair is 
$\sigma_{\gamma \gamma} n_{\gamma} l \sim 10^{-6}$, where $l \sim 10~\micron$ is the length of the forward-moving photon cloud. The total number of backward-photons is $n_{\gamma} S L \sim 10^{13}$, where $L \approx 70~\micron$ is the laser depletion length and $S \approx 25~\micron^2$ is the cross-section of the cloud, assuming the length in the third dimension is the laser spot diameter. The predicted pair yield is $10^7$, which matches the yield evaluated using the developed module for the linear BW process~\textcolor{black}{\cite{Hubbell_jpcrfd_1980}}. \RC{A similar module implemented by us into the PIC code EPOCH~\cite{Epoch} that has a different approach for treating emitted photons produced a comparable yield.} A time integrated density of the pair-production events is shown in Fig.\,2(e).

\begin{figure}[htb]
 \centering
 \includegraphics[width=\linewidth]{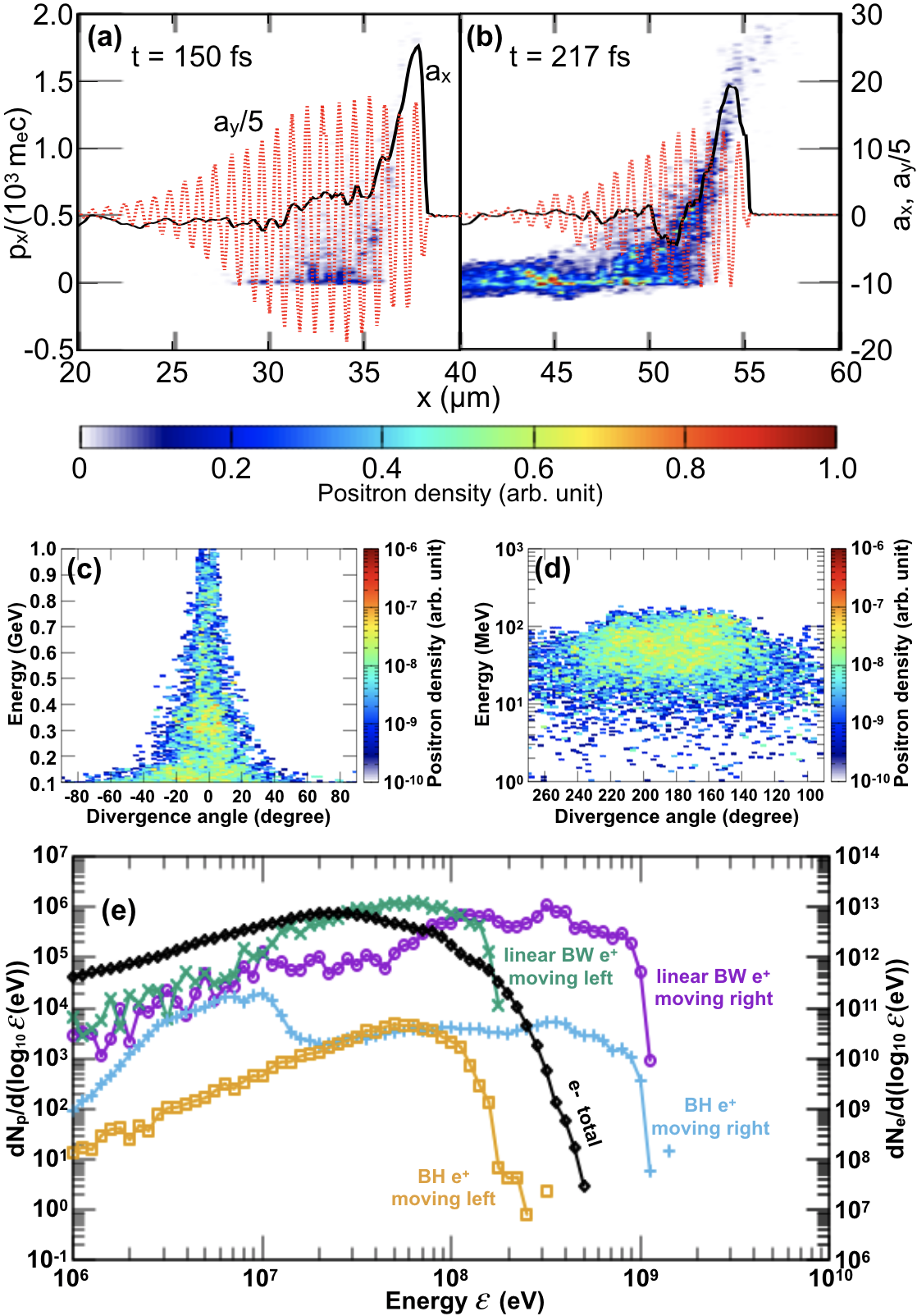}
 \caption{Laser-driven positron accelerator. (a)\&(b) Positron distribution in ($x$,$p_x$) space for $|y|\leq 2~\micron$ and electric field profiles at $t=150$\,fs and 217\,fs. The fields are averaged over $|y|\leq 2~\micron$. (c)\&(d) Energy vs. divergence angle of forward- and backward-moving positrons at $t=317$\,fs. (e) Energy spectra at $t=317$\,fs of positrons produced via the linear BW process and via the BH process, and electrons. \RC{The 3rd dimension is set to $5~\micron$ to evaluate number of particles.}
 }
 \label{fig3}
\end{figure}

\RC{The $\gamma$-ray collider is moving with the laser, continuously producing positrons with a mildly relativistic momentum $p \sim m_e c$ within the laser pulse [see Fig.\ \ref{fig3}(a)]. The positron dynamics is strongly influenced by the laser and plasma fields,  with two distinct populations emerging over time: forward-moving positrons whose energies reach 1~GeV and backward-moving positrons whose energies reach 100~MeV. Figures~\ref{fig3}(c)\&(d) show terminal positron distributions in the energy-angle space for the forward- and backward positrons. Figure~\ref{fig3}(e) shows the electron and positron energy spectra, distinguishing the linear BW and Bethe-Heitler [see Supplemental Material] positrons to emphasize the dominant role of the linear BW process. A striking feature of Fig.~\ref{fig3}(e) is that the peak energy of forward positrons exceeds the peak energy of forward electrons by a factor of two. The electrons gain their energy from the laser via the direct laser acceleration assisted by the plasma magnetic field~\cite{gong.PRE.2020}, but the positrons are not able to do that because they are positively charged. The plasma magnetic field deflects positrons backward rather than forward, which causes the formation of the backward positron population.}



\RC{We tracked the energetic forward-moving positrons and found that they gain most of their energy (80\%) from the strong forward-moving longitudinal plasma electric field, thus discovering a new positron acceleration mechanism. Figure~\ref{fig3}(b) confirms that the energetic positrons are surfing with the spike in $a_x$. The positrons continue accelerating until they overtake the laser pulse or leave the acceleration region in lateral direction. The acceleration by $a_x$ only works for positrons, whereas the same field pulls plasma electrons backward creating the backward emission that contributes to the photon collider.}

\RC{The discovered acceleration mechanism produces $10^6$ or $0.1$~pC of positrons with energies above 100~MeV and  average divergence angle $|\theta| \sim 10^{\circ}$. The high plasma density is not only important for generating strong $a_x$ needed for positron acceleration (no $a_x$ spike is produced at subcritical densities~\cite{martinez.prab.2023}), but it is also crucial for achieving a high number of accelerated positrons. Positrons must catch up with $a_x$ to experience the acceleration, but this is hard to achieve if $a_x$, whose speed is $u$, moves too fast. In a low density plasma, $u$ is close to the group velocity $v_g/c \approx \sqrt{1-n_e/\gamma_L n_c}$~\cite{Sentoku_fst_2006}. In a dense plasma, $u$ is lower than $v_g$ due to laser depletion, which enables more positrons to experience acceleration. In our case, $u/c \approx 0.8$, but $v_g/c \approx 0.98$. Only relativistic positrons with $v_x \approx c \cos \theta > u$ are able to catch up with $a_x$. We have $v_x \approx c \cos \theta > u$ for $|\theta| \leq 37^{\circ} $, whereas $v_x \approx c \cos \theta > v_g$ for $|\theta| \leq 11^{\circ} $. The 20\% reduction in $u$ compared to $v_g$ increases the range of $\theta$ by a factor of three and thus significantly increase the number of positrons that can catch up with $a_x$.}

\begin{figure}[t!]
 \centering
 \includegraphics[width=\linewidth]{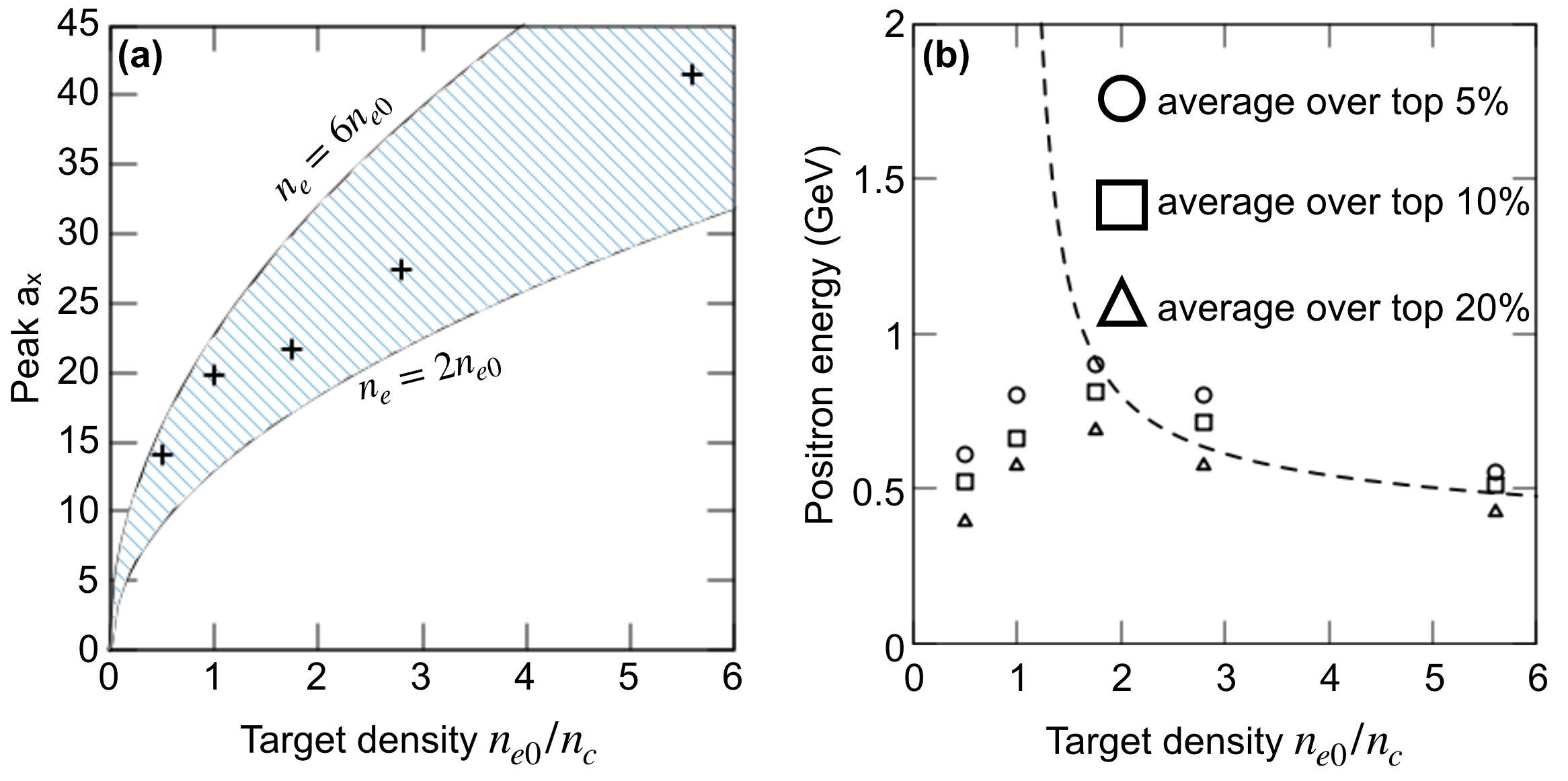}
 \caption{(a) Normalized electric field $a_x$ at the leading edge of the laser pulse as a function of target density. The shaded area is given by Eq.~(1) for $2n_{e0} \le n_e \le 6n_{e0}$, $a_L = 120$, and $u=0.8c$. (b) Positron energies averaged over the top 5\%, 10\%, and 20\% of the positron spectra for different target densities. The dotted curve is $\Delta \epsilon_{e^{+}} = c\Delta p_{e^{+}}$ obtained from Eq.~(2) for $a_L = 120$, $u=0.8c$, and $n_e = 4n_{e0}$.}
 \label{fig5}
\end{figure}

To examine the impact of the plasma density $n_{e0}$ on the strength of $a_x$ and the positron energy gain, we performed extra simulations with $n_{e0} / n_c =$ 0.5, 1.0, 1.75, and 5.6. Figures~\ref{fig5}(a)\&(b) show $a_x$ at the leading edge of the pulse and the energy gain by forward-moving positrons versus $n_{e0}$. We average $a_x$ over $y$ at the time when the laser peak intensity reaches the pulse leading edge to obtain the values in Fig.~\ref{fig5}(a). The energies in Fig.~\ref{fig5}(b) were averaged over the top 5, 10, and 20 percent of the positron spectrum to confirm the trend. The discovered regime is robust and can be achieved over a wide range of plasma densities. For $n_c \le n_{e0} \le 5.6\,n_c$, the number of positron with energies above 100\,MeV and $|\theta| \lesssim 10^\circ$ is consistently about $10^6$. At $n_{e0} / n_c = 0.5$, the speed of $a_x$ is very close to $c$, which makes $a_x$ too fast to effectively accelerate positrons that are originally only mildly relativistic. 

We next use estimates for $a_x$ and the positron energy gain to determine their scaling at high $n_{e0}$. The electron density pileup responsible for $a_x$ is sustained due to force balance, $0=F_p + F_s$, between the laser ponderomotive force $F_p = -m_e c^{2}\nabla_x \gamma_L$ and $F_s= - a_x m_e c \omega_0$. We estimate that $\gamma_L / |\nabla_x \gamma_L | \simeq l_{\rm skin}$, where $l_{\rm skin}=\sqrt{\gamma_L} c/\omega_{pe}$ is the relativistic skin depth. Taking into account that $a_L \gg 1$, we obtain
\begin{equation}
	\label{eq1}
	a_{x} \simeq \sqrt{\gamma_L n_e / n_c},
\end{equation}
where $n_e$ is the density of the electron pileup. The shaded area in Fig.\ \ref{fig5}(a) shows $a_x$ from Eq.\,\eqref{eq1} for $a_L=120$ and $2n_{e0} \le n_e \le 6n_{e0}$. The latter is the entire range of $n_e$ observed in the simulations, with $n_e \approx 2n_{e0}$ for $n_{e0}=5.6 n_c$ and $n_e \approx 6n_{e0}$ for $n_{e0}= 0.5n_c$. The momentum gain, $\Delta p_{e^{+}}$, from $a_x$ can be estimated by integrating the positron equation of motion $dp_{e^{+}}/dt \simeq m_ec \omega_0 {\bar a}_x$ over the acceleration time interval $\Delta t_{\rm acc}$, where ${\bar a}_x=a_x/2$ is the average field amplitude in the acceleration region. The length of the region with positive $a_x$ is the width of the electron pileup, $l_{\rm skin}$, plus the length of the positively-charged electron cavity, $l_{\rm cav}$, formed behind the pulse leading edge. We estimate $l_{\rm cav}$ from the charge conservation: $\left( n_{e0}-n_{c} \right) l_{\rm cav} = \left( n_e-n_{e0} \right) l_{\rm skin}$ for $n_{e0}>n_c$. The acceleration region is moving forward with velocity $u$ while the positron velocity is $v_x$, so that $\Delta t_{\rm acc} \equiv (l_{\rm cav} + l_{\rm skin})/(v_x-u)$. Assuming an ultra-relativistic positron, we set $v_x \sim c$. After taking into account that $\gamma_L n_{c} \gg n_e$ for $a_L \gg 1$, we find that that the positron momentum gain is 
	\begin{equation}
	\label{eq4}
	 \Delta p_{e^{+}} \simeq 
 	\frac{\gamma_L m_ec}{2}\frac{1}{1-u/c} \ 
	 \frac{n_{e} - n_c}{n_{e0} - n_c}.
	\end{equation}
Equation~\eqref{eq4} gives $\Delta p_{e^{+}}/m_ec\simeq 1200$ for $n_{e0}=2.8\,n_c$, $a_L = 120$, $u=0.8c$, and $n_e=4n_{e0}$, reproducing the significant positron momentum increase at the pulse leading edge seen in Fig.\,\ref{fig3}(b). The energy gain, $\Delta \epsilon_{e^{+}} = c\Delta p_{e^{+}}$, obtained from Eq.~\eqref{eq4} is shown in Fig.\,\ref{fig5}(b) with a dashed curve. For high densities, $\Delta \epsilon_{e^{+}}$ has a weak dependence on $n_{e0}$, because the increase in $a_x$ is counteracted by the reduction in the acceleration time caused by lower $u$.



\RC{In summary, we discovered a robust regime where a laser-irradiated plasma self-organizes to produce positrons and accelerate them. The GeV-level positron beam can be generated using just a single laser with an experimentally available intensity. The regime requires the use of a dense plasma that can create a strong longitudinal electric field via electron pileup. The field is crucial for creating the $\gamma$-ray collider and for accelerating positrons. The positron acceleration was discovered by a first-of-its-kind simulation code generating pairs via photon-photon collisions. This code has direct relevance to astrophysics research since correct treatment of secondary pairs is one of the main problems facing modern PIC simulations of pulsars~\cite{Philippov_2018,Hakobyan_2023} The uniform density is a simplification and not a requirement. A simulation with $n_e$ ramping up from 0.5 to $3 n_c$ over $60~\micron$ has a similar pair yield of $10^7$. 3D simulations with PICLS (see Supplemental Materials) and EPOCH~\cite{Epoch} have $n_{\gamma}$ that is similar to that in our 2D simulations, confirming the robustness of the discussed phenomena.} Lastly, our regime can be instrumental in gauging the focal intensity of multi-PW lasers. At $10^{21}$\,W/cm$^2$, the positron yield is five orders of magnitude lower than at $10^{22}$\,W/cm$^2$. Therefore, the presence of energetic positrons in the laser direction can be a confirmation of laser intensity exceeding $10^{22}$\,W/cm$^2$.


This study was supported by JSPS KAKENHI Grants No. JP19KK0072, No. JP20K14439, No. JP20H00140, No. JP22J10867, JP23K03354, and JST PRESTO Grant No. JPMJPR21O1. The work by Y.H., I.-L. Y., K. T., and A. A. was supported by AFOSR (Grant No. FA9550-17-1-0382) and by National Science Foundation  – Czech Science Foundation partnership (NSF award PHY-2206777). \\

\bibliographystyle{apsrev4-1}
\bibliography{reference}

\end{document}